\documentclass[aps,physrev,twocolumn,showpacs,superscriptaddress,amsmath,amssymb,nofootinbib]{revtex4-2}

\usepackage{amsmath}
\usepackage{graphicx}
\usepackage{appendix}
\usepackage{adjustbox}
\usepackage{bm}
\usepackage{ulem}
\usepackage[colorlinks=true, allcolors=blue]{hyperref}
\usepackage{orcidlink}

\newcommand{\moe}{\affiliation{Key Laboratory of Atomic and Subatomic Structure and Quantum Control (MOE), Guangdong-Hong Kong Joint Laboratory of Quantum Matter, Guangzhou 510006, China
}}

\newcommand{\sfim}{\affiliation{Guangdong Basic Research Center of Excellence for Structure and Fundamental Interactions of Matter, Guangdong Provincial Key Laboratory of Nuclear Science, Guangzhou 510006, China}}

\newcommand{\iqm}{\affiliation{State Key Laboratory of Nuclear Physics and Technology, Institute of Quantum Matter, South China Normal University, Guangzhou 510006, China}}

\newcommand{\scnt}{\affiliation{Southern Center for Nuclear-Science Theory (SCNT), Institute of Modern Physics, Chinese Academy of Sciences, Huizhou 516000, Guangdong Province, China}}

\newcommand{\gscas}{\affiliation{Graduate School of China Academy of Engineering Physics, Beijing 100193, China}}

\newcommand{\peku}{\affiliation{School of Physics, Peking University, Beijing 100871, China}}

\begin{document}

\title{Dilated coordinate method for solving nuclear lattice effective field theory}
\iqm
\gscas

\author{Guangzhao He\orcidlink{0009-0007-9418-1840}}
\iqm
\moe
\sfim
\author{Zhenyu Zhang\orcidlink{0000-0001-7042-2311}}
\iqm
\moe
\sfim
\author{Teng Wang}
\peku
\author{Qian Wang\orcidlink{0000-0002-2447-111X}}
\email{qianwang@m.scnu.edu.cn}
\iqm
\sfim
\scnt
\author{Bing-Nan Lu}
\email{bnlv@gscaep.ac.cn}
\gscas

\date{\today}

\begin{abstract}

We introduce a dilated coordinate method to address computational challenges in nuclear lattice effective field theory (NLEFT) for weakly-bound few-body systems. The approach employs adaptive mesh refinement via analytic coordinate transformations, dynamically adjusting spatial resolution to resolve short-range nuclear interactions with fine grids while efficiently capturing long-range wave function tails with coarse grids.
Numerical demonstrations for two- and three-body systems confirm accelerated convergence towards infinite-volume limit compared to uniform lattices, particularly beneficial for accessing highly excited states and shallow bound states near the continuum threshold. 
This method establishes a foundation for \textit{ab initio} studies of exotic nuclear systems near the dripline and light hypernuclei, with direct extensions to scattering and reaction processes.

\end{abstract}
\maketitle

\section{Introduction}

Effective field theory (EFT) serves as the standard paradigm for describing systems exhibiting explicit scale separation~\cite{Ann.Rev.Nucl.Part.Sci.43-209, IJMPA31-1630007}.
Each EFT possesses an intrinsic momentum scale $\Lambda$ that separates the soft scale, where the important low-energy physics resides, and the hard scale containing the irrelevant high-energy degrees of freedom. 
In a typical EFT, high-energy degrees of freedom are integrated out and represented by short-range structureless interactions arranged according to a specific power-counting scheme.
The basic assumption of the EFT is that the low-energy physics is independent of $\Lambda$ within an uncertainty scaling as $O((Q/\Lambda)^N)$, where $Q$ denotes the characteristic momentum of soft modes and $N$ indicates the order of the interaction.
Generally, we require $Q \ll \Lambda$ for a fast convergence of the results.

Nuclear physics provides a natural testing ground for EFT concepts. 
Although quantum chromodynamics (QCD) fundamentally governs the nuclear structure, quarks and gluons become confined within hadrons during low-energy processes that are fully dictated by the corresponding EFT. 
Over recent decades, substantial efforts have focused on constructing nucleon-nucleon interactions rooted in fundamental symmetries within the EFT framework~\cite{RMP81-1773, Phys.Rept.503-1, RMP92-025004, Front.Phys.8-98,  PPNP137-104117}. 
Resulting chiral nuclear forces now enable extensive first-principles investigations of low-energy nuclear structure and reactions~\cite{RMP87-1067, PPNP69-131, Rep.Prog.Phys.77-096302, Phys.Rept.621-165, PPNP63-117}. 
Here the soft scale corresponds to the nucleon separation momentum $Q \sim \sqrt{ME} \approx 100$~MeV, where $M$ represents nucleon mass and $E$ denotes typical single-nucleon separation energy. Practitioners typically adopt $\Lambda \approx 500$~MeV for the EFT momentum cutoff. 
Additional soft scales exist, including the pion mass $M_{\pi} \approx 140$~MeV governing long-range components of the interaction~\cite{PRC68-041001, NPA747-362, PRL128-142002}.

An efficient method to regularize the EFT is utilizing a spatial lattice.
The lattice spacing $a$ establishes an anisotropic momentum cutoff $\Lambda \approx \pi/a$.
The resulting regularized nuclear force can be integrated into the many-body Schr\"odinger equation, enabling solutions via sparse matrix algebra or quantum Monte Carlo techniques within the framework of nuclear lattice effective field theory (NLEFT)~\cite{PRC70-014007, PRC72-024006, PPNP63-117}.
This \textit{ab initio} methodology has successfully facilitated extensive studies of nuclear ground states~\cite{g1,g2,g3,g4,g5,g6,g7,pn99-6dxt}, excited states~\cite{ex1,ex2,ex3,ex4}, scattering phenomena~\cite{st1,st2}, nucleon thermodynamics~\cite{ft1,ft2,ft3,ft4}, clustering effects~\cite{clust1,clust2,clust3,clust4,clust5} and hypernuclei~\cite{EPJA56-248, EPJA58-167, EPJA60-215, Sci.Bull.70-825}. 
Recent investigations further extend NLEFT applications from nucleonic to hadronic systems~\cite{Zhangz:2024yfj}.
Typical NLEFT implementations employ lattice spacings spanning $1$~fm to $2$~fm, corresponding to momentum cutoffs ranging between $300$~MeV and $600$~MeV.

Unlike the case of solids in condensed matter physics, where the lattice is composed of real ions, in NLEFT the lattice is artificially introduced, and the final results should not depend on the lattice spacing and the specific lattice geometry.
Comparing the results obtained with different lattice spacings and geometries can be used to constrain the size of lattice artifacts.
However, solving quantum many-body systems on non-standard lattices poses substantial computational difficulties.
Consequently, most NLEFT calculations are constrained to single-spacing cubic lattices.  
Only recently was the body-centered cubic (BCC) lattice introduced into NLEFT and applied to calculate the Bertsch parameter for the unitary Fermi gas, yielding excellent agreement with standard cubic lattice~\cite{PRA101-063615, PRC104-044304}.
The central challenge involves implementing kinetic energy operators through finite differencing among adjacent BCC lattice points.

The development of non-standard lattice algorithms would also be beneficial for investigating shallow nuclear bound states, as well as processes involving low-energy scattering or reactions.
For instance, the deuteron binding energy of $-2.224$~MeV and the triton ($^3$H) binding energy of $-8.482$~MeV are both significantly smaller than the resolution scales typically employed in EFTs.
The $\Lambda$-hypernucleus $^3_{\Lambda}$H exhibits an even smaller $\Lambda$-separation energy of approximately $0.102$~MeV~\cite{PRL131-102302}, resulting in a spatially extended $\Lambda$ wave function that necessitates extraordinarily large simulation volumes in lattice calculations~\cite{EPJA60-215}.
Similarly, halo nuclei or clustered states such as the Hoyle state consist of weakly bounded nuclear fragments, motivating the development of halo/cluster EFTs~\cite{NPA712-37, PLB569-159}.
Finally, stellar nuclear reactions typically occur below $1$~MeV~\cite{PPNP54-535, PPNP89-56}.
Such low-energy bound states and 
reactions exhibit high sensitivity to boundary conditions and lattice artifacts, requiring simultaneously a small lattice spacing and a large volume to achieve the desired accuracies. 
In such scenarios, the standard uniform cubic lattice proves particularly inefficient.

In this work, we further exploit the freedom of choosing lattice spacing and geometry, exploring the usage of non-uniform lattices for solving few-body systems.
Such adaptive mesh refinement (AMR) technique employs variable lattice spacing adapted to the local physical characteristics, significantly reducing computational complexity in multiscale simulations. 
Generally, finer resolutions apply where physical properties exhibit rapid spatial variation, while coarser grids suffice otherwise.  
This methodology has found broad application across multiple domains, spanning numerical relativity~\cite{enzo,grchombo}, chemical system modeling~\cite{chem1}, magnetic field computations~\cite{meg1,meg2,meg3}, materials simulation~\cite{mat}, mechanical analysis~\cite{f,f2}, computational fluid dynamics~\cite{cfd1,cfd2} and differential equation solutions~\cite{cal1,cal2}.
Lattice QCD utilizes adaptive multigrid algorithms for solving Dirac operator~\cite{PhysRevLett.100.041601, PhysRevLett.105.201602}.  
Despite such demonstrated efficacy, so far adaptive mesh has not been applied in solving nuclear EFT, which exhibits inherent multiscale nature compatible with the philosophy of the AMR methods.

As the first step of linking the AMR technique to the nuclear EFT, here we focus on the few-body system ($A \leq 3$) accessible through direct diagonalization methods such as Lanczos algorithm~\cite{NPA768-179, EPJA41-125}. 
We examine systems governed by strong short-range attractive interactions emulating nuclear forces, solving the corresponding Schr\"odinger equation using a specialized AMR method called dilated coordinate method.  
This methodology is particularly relevant for light nuclei possessing shallow bound states.
Accurately capturing highly oscillatory wave functions within interaction ranges demands small lattice spacings $a \approx \pi / \Lambda$, whereas the smoother asymptotic tails outside these regions permit larger spacings $a \approx \pi / Q$.  
The dilated coordinate method implements locally refined meshes in regions of nucleon proximity and employs coarser grids beyond the interaction ranges.
Given the characteristic scale separation $Q \ll \Lambda$ inherent in nuclear EFT, this strategy significantly reduces the number of lattice points required to minimize finite volume effects, offering particular advantages for shallow bound systems.
As a demonstration of principles, this work presents the method and provides numerical benchmarks using several toy models.  
The application to realistic short-range nuclear forces is straightforward and reserved for future works.

This paper is organized as follows. 
Following this introduction, we develop the dilated coordinate method as an extension of the adaptive mesh approach and present numerical results for two- and three-nucleon systems across one- and three-dimensional configurations. 
We conclude by summarizing the findings and discussing potential applications within future NLEFT studies.

\section{Dilated coordinate method}
\subsection{General consideration}
\label{sec:1}
At leading order in the EFT expansion, the nuclear force is represented by an attractive potential with a short range of approximately $1$~fm.
This interaction must be sufficiently strong to support bound states like the deuteron and triton in few-body systems.
To solve the corresponding two- and three-body Schr\"odinger equations, we employ direct diagonalization via the Lanczos algorithm.
Working in the center-of-mass frame, we represent the wave functions using relative coordinates.
For compatibility with lattice calculations, we utilize the relative displacements instead of Jacobi coordinates.
Specifically, for a two-body system, we take the relative coordinate $\bm{r} = \bm{r}_1 - \bm{r}_2$.
For a three-body system, we use $\bm{r}_{13} = \bm{r}_1 - \bm{r}_3$ and $\bm{r}_{23} = \bm{r}_2 - \bm{r}_3$ as the lattice coordinates.
Accordingly, the Schr\"odinger equations become differential equations of these coordinates.

\begin{figure}[h]
\centering
\includegraphics[width=0.4\textwidth]{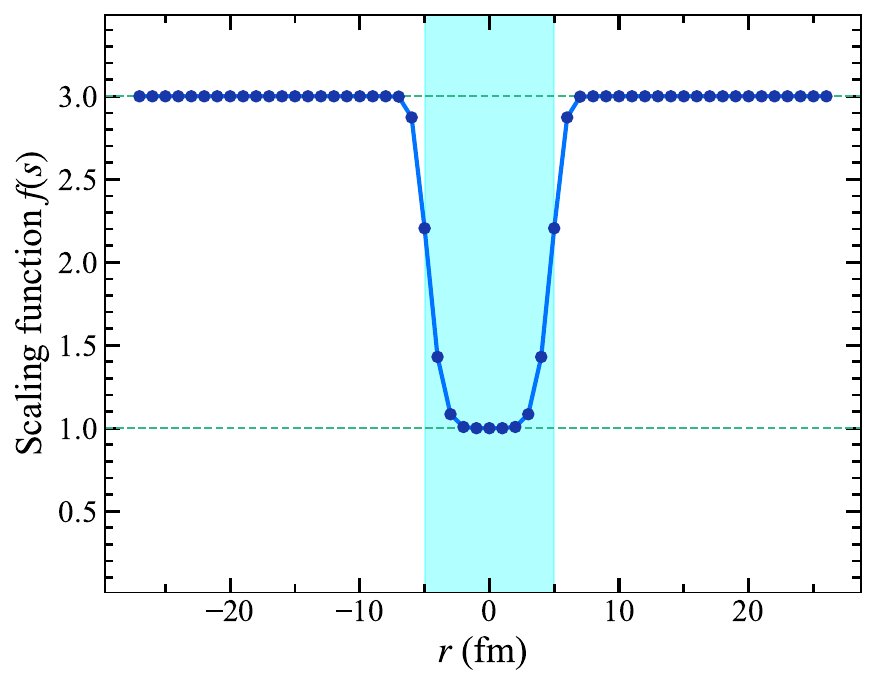}
\caption{Example of the scaling function $f(s)$ with radius $R = 5$~fm and asymptotic ratio  $\lambda=3$. Blue shadow denotes the range $r \leq R$ where the lattice spacing is unchanged.}
\label{fig:1}       
\end{figure}

For shallowly bound states, the wave functions exhibit rapid oscillations within the region where two nucleons are close to each other, while varying slowly outside this region. Denoting the interaction range as $R$, this implies that for a two-body system, a dense spatial mesh is required for $r \leq R$, while a sparser mesh suffices for $r > R$. To implement such a non-uniform mesh, we introduce an auxiliary vector $\bm{s}$ and define a transformation
\begin{equation}
    \bm{r} = \bm{s} f(s),
    \label{eq:mapping}
\end{equation}
where $s = |\bm{s}|$ and $f(s)$ is a scaling function:
\begin{equation}
    f(s) = 1 + (\lambda - 1) \left[ 1 - \exp \left( -s^{6} / R^{6} \right) \right],
    \label{eq:dialteFunction}
\end{equation}
with $\lambda > 1$ and $R$ tunable parameters. This function satisfies $f(s) \approx 1$ for $|s| \leq R$ and asymptotically approaches $\lambda$ for $|s| \gg R$, which is exemplified in Fig.~\ref{fig:1} with $\lambda=3$ and $R=5$ fm. 

\begin{figure}[h]
\centering
\includegraphics[width=0.37\textwidth]{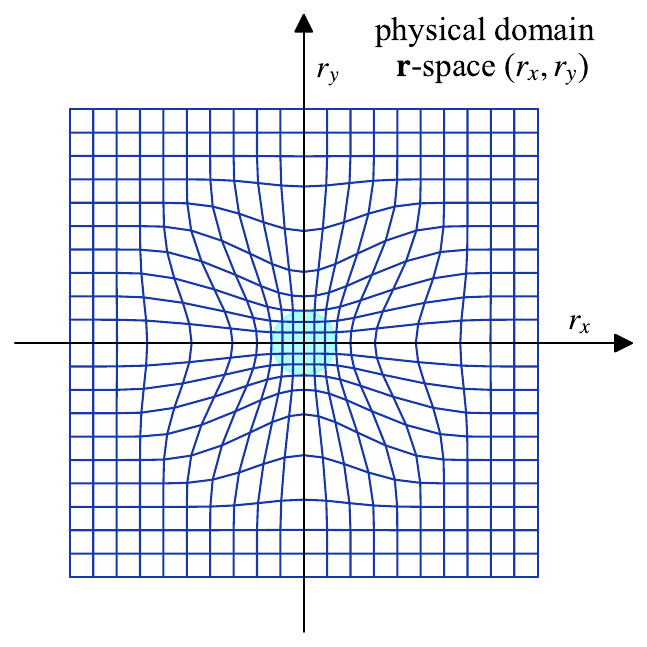}
\includegraphics[width=0.37\textwidth]{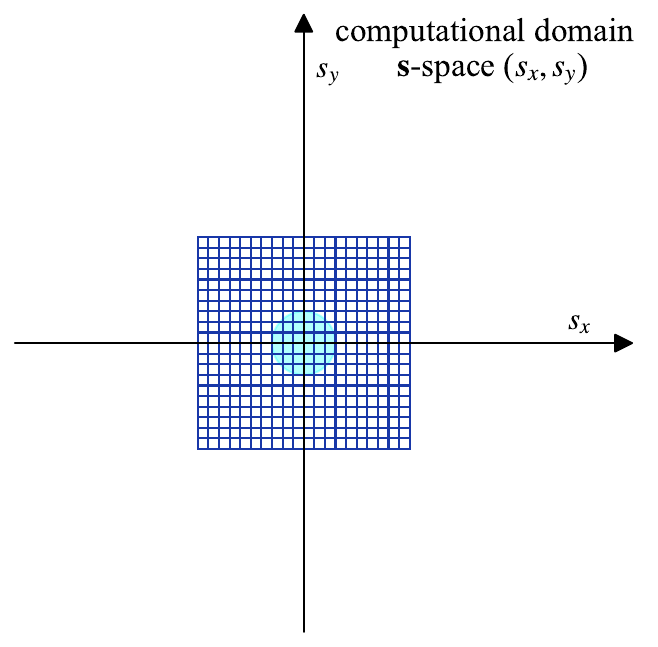}
\caption{Schematic representations of the dilated coordinate in physical $\bm{r}$-space and the uniform cubic lattice in $\bm{s}$-space. The blue circles indicate compact domain $r \leq R$.
}
\label{fig:1-1}       
\end{figure}

The transformation Eq.~(\ref{eq:mapping}) maps a standard cubic lattice (computational $\bm{s}$-space) into a non-uniform lattice (physical $\bm{r}$-space). Importantly, the mapping is almost trivial for $|\bm{s}| \leq R$ and approaches $\bm{r} = \lambda \bm{s}$ for $|\bm{s}| \gg R$. Fig.~\ref{fig:1-1} schematically illustrates these features for two dimensions, where the blue circle in the center represents the region nearly unchanged under the transformation.
Since the computational $\bm{s}$-space is an equidistant grid, operations such as differentiations and integrations can be normally performed on the $\bm{s}$-grid using mature algorithms like finite difference.
However, the real physics is represented in the dilated $\bm{r}$-coordinates.
We should transform the Schr\"odinger equation from $\bm{r}$- to $\bm{s}$-space and solve it numerically.
Note that the dilation transformation introduced here is conceptually similar to standard coordinate transformations (\textit{e.g.}, Cartesian to spherical), where the real physics is independent of the coordinate system employed to describe it.

\subsection{Two-body system}

For a two-body system, the Schr\"odinger equation in the center-of-mass frame is given by
\begin{align}
    \left[-\frac{1}{2 \mu} \nabla ^2 + V(\bm{r})\right] \psi(\bm{r})=E \psi(\bm{r}),
\end{align}
where $\mu = M/2$ is the reduced mass.
After applying the transformation Eq.~(\ref{eq:mapping}), we obtain the equation for $\bm{s}$:
\begin{align}
    \left[-\frac{1}{2 \mu} \sum_i \left( \sum_j \frac{\partial s_j}{\partial r_i} \frac{\partial}{\partial s_j} \right)^2 +\widetilde{V}(\bm{s}) \right] \widetilde{\psi}(\bm{s})=E \widetilde{\psi}(\bm{s}),
    \label{eq:2NSchrodinger}
\end{align}
with the abbreviations $\widetilde{V}(\bm{s}):=V(\bm{s} f(s))$ and $\widetilde{\psi}(\bm{s}):=\psi(\bm{s} f(s))$.
The $3\times 3$ Jacobi matrix in Eq.~(\ref{eq:2NSchrodinger}) can be evaluated via matrix inversion,
\begin{eqnarray}
\left[\frac{\partial \bm{s}}{\partial \bm{r}}\right]=\left[\frac{\partial \bm{r}}{\partial \bm{s}}\right]^{-1},
\end{eqnarray}
where the matrix elements on the right-hand side are given analytically by Eq.~(\ref{eq:dialteFunction}):
\begin{eqnarray}
    \frac{\partial r_i}{\partial s_j} = \delta_{ij} f(s) + \frac{s_is_j}{s} f'(s).
\end{eqnarray}
The partial derivatives with respect to $\bm{s}$ in Eq.~(\ref{eq:2NSchrodinger}) can be evaluated using the finite difference or FFT methods.
This approach allows solving the transformed Schr\"odinger equation Eq.~(\ref{eq:2NSchrodinger}) on a uniform lattice in $\bm{s}$-space, which is equivalent to solving the original Schr\"odinger equation on an adaptive mesh with variable lattice spacing. 
Therefore, we can yield the same eigenenergies base on dilated coordinate.
For observables beyond energies, the corresponding operators must be transformed accordingly.

\subsection{Three-body system}

For a three-body system with identical mass $M$, we employ two relative coordinates $\bm{r}_{13}$ and $\bm{r}_{23}$.
The corresponding Schr\"odinger equation is
\begin{equation}
    \left[ - \frac{1}{M}(\nabla_{13}^2+\nabla_{23}^2+\nabla_{13} \cdot \nabla_{23}) + V(\overline{\bm{r}}) \right] \psi(\overline{\bm{r}}) = E \psi(\overline{\bm{r}}),
    \label{eq:3NScrhodinger}
\end{equation}
where $\overline{\bm{r}} = {\bm{r}_{13}, \bm{r}_{23}}$ denotes the set of relative coordinates, $\nabla_{13}$ and $\nabla_{23}$ represent gradients with respect to $\bm{r}_{13}$ and $\bm{r}_{23}$, respectively.
The center-of-mass degree of freedom has been separated and removed.
The potential $V(\overline{\bm{r}})$ includes two- and three-body components, both short-ranged and sufficiently strong to support two- and three-body bound states.
This scenario relates closely to three-nucleon system calculations within NLEFT~\cite{NPA768-179, EPJA41-125}, though spin and isospin degrees of freedom are omitted as they are not essential.
Accurate determination of three-body forces in NLEFT requires efficient solutions for light nuclei like $^3$H, which has been constrained by the finite volume effects. 
The dimension of the Hilbert space increases rapidly with lattice size, generally necessitating extrapolations~\cite{EPJA54-121, EPJA60-215}.
The dilated coordinate method offers a potential solution.

\begin{figure}[ht!]
\centering
\includegraphics[width=0.5\textwidth]{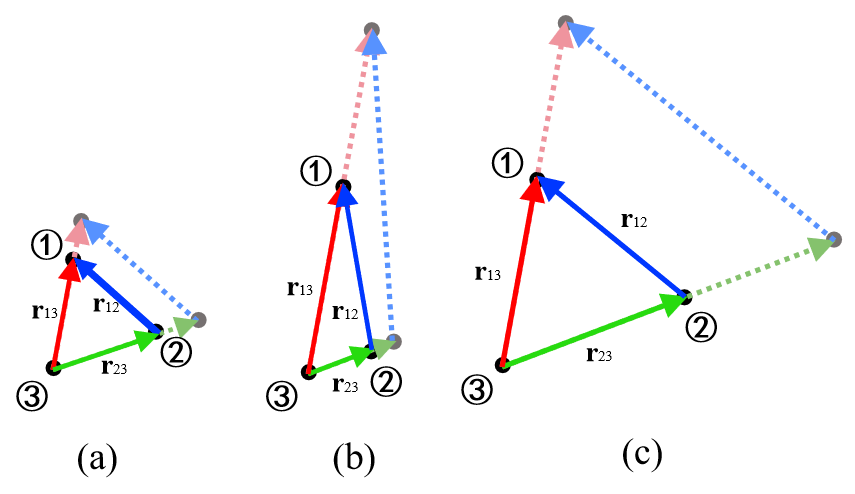}
\caption{Schematic plot of relative displacements between three particles. 
(a) Three-body compact configuration; (b) $2 + 1$ molecular configuration; (3) $1 + 1 + 1$ loose configuration.
Black (grey) dots and solid (dotted) vectors indicate the initial (dilated) positions.}
\label{fig:2}       
\end{figure}

To implement the dilated coordinate method, we apply the transformation
\begin{equation}
    \bm{r}_{13}=\bm{s}_{13} f(s_{13}), \quad 
    \bm{r}_{23}=\bm{s}_{23} f(s_{23}),
    \label{eq:3Ntransformation}
\end{equation}
with relative distances $s_{13} = |\bm{s}_{13}|$ and $s_{23} = |\bm{s}_{23}|$,  and $f(s)$ is identical to the two-body case in Eq.~(\ref{eq:dialteFunction}).
The kinetic energy term in Eq.~(\ref{eq:3NScrhodinger}) can be reformulated using partial derivatives with respect to $\bm{s}$, following the approach in Eq.~(\ref{eq:2NSchrodinger}).
The computation of the Jacobi matrix and the implementation of lattice derivatives via finite difference or FFT parallel the two-body methodology.
Fig.~\ref{fig:2} schematically illustrates the three-nucleon configurations before and after dilation. 

For identical bosonic three-body systems, exchange symmetry plays a crucial role, which means that the exchange of any particle pair leaves the wave function and physical outcomes unchanged.
Consequently, the three-body wave function must be symmetrized,
\begin{eqnarray}
\widetilde{\psi}_{\rm symm} (\bm{s}_{13}, \bm{s}_{23}) &=& \frac{1}{6}\left(1 + P_{12} + P_{23} + P_{13} \right.
 \nonumber \\
  && \left. +P_{12}P_{23}+P_{12}P_{13} \right)  \widetilde{\psi}(\bm{s}_{13}, \bm{s}_{23})
  \label{eq:symmetrization}
\end{eqnarray}
where $P_{ij}$ denotes the exchange operator for particles $i$ and $j$. 
We exchange the $\bm{s}$-coordinates rather than the $\bm{r}$-coordinates. 
This approach allows direct application of coordinate exchanges on the uniform $\bm{s}$-lattice. 
For particles with spin and isospin degrees of freedom, these quantum numbers must also be exchanged during the operation.
This symmetrization ensures all three particles experience the dilation transformation equally, removing any bias associated with particle labeling.
Critically, it symmetrizes the biased transformation described by Eq.~(\ref{eq:3Ntransformation}).
We observe that this operation significantly reduces lattice artifacts.
Particularly, in some cases the dilated lattice results converge to incorrect infinite-volume limit without the wave function symmetrization.

\section{Results and Discussions}
\subsection{One-dimensional two-boson system}

In what follows we numerically benchmark the dilated coordinate method using various interaction models in one- and three-dimensions. We first consider a one-dimensional toy model with two identical spinless bosons interacting via a short-range Gaussian potential:
\begin{equation}
V(x) = C \exp\left(-\frac{r^2}{2\sigma^2}\right),
\label{eq:GaussianPot}
\end{equation}
with $C=-1.5$ MeV the interaction strength and $\sigma=2.0$ fm the interaction range. The mass is taken as $M=938.92$ MeV, and we focus on the ground state energy. 
Here we have adjusted the strength of the interaction so that this potential has a single shallow bound state.
\begin{figure}[h]
\centering
\includegraphics[width=0.45\textwidth]{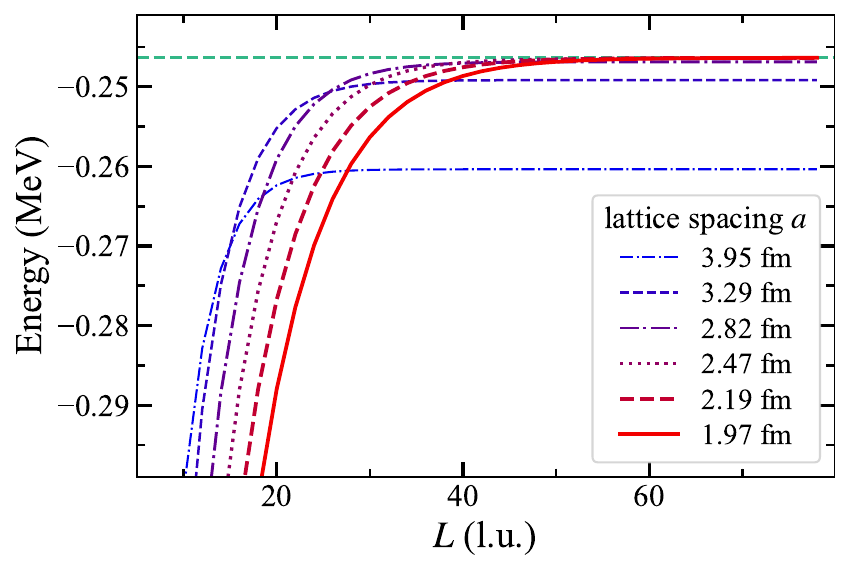}
\caption{Binding energies versus box size $L$ for a one-dimensional two-body system computed with with different lattice spacing $a$ in uniform lattices. The horizontal dashed line indicates the simultaneously infinite-volume ($L\rightarrow \infty$) and continuum ($a\rightarrow 0$) limit.}
\label{fig:cutoff}
\end{figure}
We first examine the dependence of the binding energy on the lattice spacing.
Fig.~{\ref{fig:cutoff}} presents the infinite volume extrapolation of the result for varying lattice spacing $2$~fm$\leq a \leq 4$~fm.
For each value of $a$, the energy converge exponentially as $L$ goes to the infinity.
The converged values depend on the lattice spacing and coincide for small $a$.
The resulting continuum limit at $a\rightarrow 0$ is $E = -0.246$~MeV, which implies a large spatial extension of the wave function.
In this case the finite volume effects are severe, requiring an extremely large volume.
Next we fix the lattice spacing to $a=0.99$~fm to completely suppress the lattice discretization errors and analyze the finite volume effects.

\begin{figure}[h]
\centering
\includegraphics[width=0.45\textwidth]{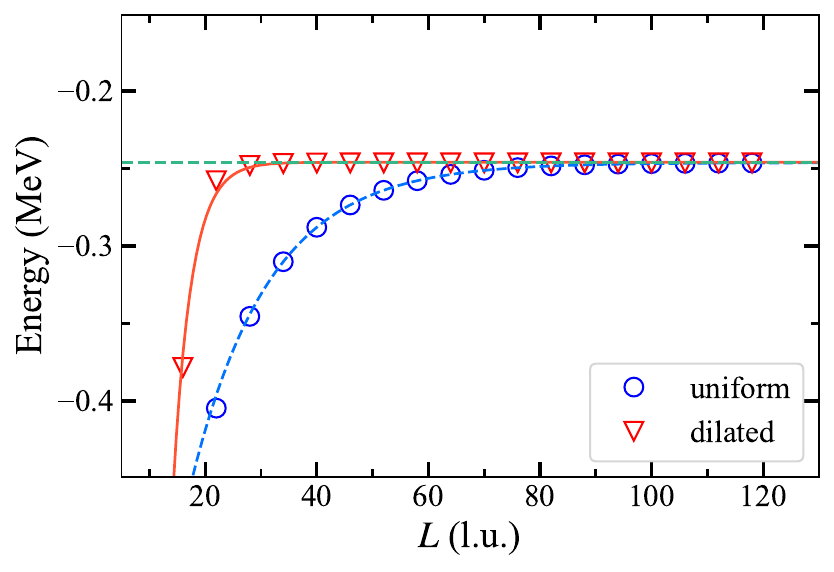}
\includegraphics[width=0.45\textwidth]{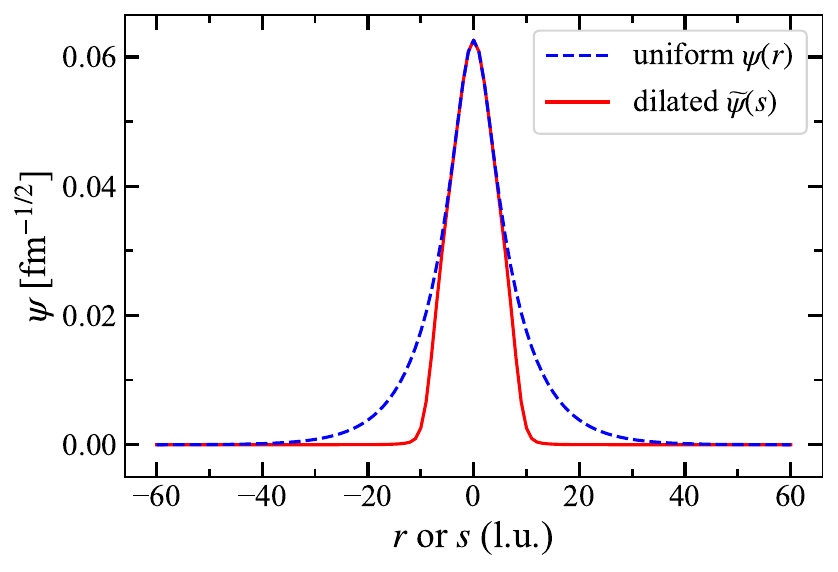}
\caption{(Upper panel) Binding energies for a one-dimensional two-body system computed with uniform (cicles) and dilated (triangles) lattices versus box size $L$.
Dilation is performed for $R = 10$~fm with the asymptotic ratio $\lambda = 3$.
(Lower panel) The same wave function against the original (dashed) and dilated coordinates (solid). 
The tail of $\tilde{\psi}(s)$ shrinks relative to $\psi (x)$ due to the dilation of the lattice at large distances.
}
\label{fig:3}
\end{figure}

The binding energies for this one-dimensional system are presented in the upper panel of Fig.~\ref{fig:3}. For comparison, we compute the energies using both a uniform lattice and a dilated lattice. The dilated lattice was generated by increasing the lattice spacing at $r > 10$~fm with an asymptotic ratio converging to $\lambda = 3$, according to the transformation in Eq.~(\ref{eq:dialteFunction}). We plot the energies as functions of the number of lattice points $L$ (equivalent to the box size in units of lattice spacing). For identical $L$, the dilated lattice describes a significantly larger volume than the uniform lattice as we have effectively increased the lattice spacing beyond the interaction range.
The solid and dashed lines represent infinite-volume extrapolations using the ansatz
\begin{align}
 &E(L) = E(\infty) + A \exp(-\kappa L), \label{eq:extrapolation}
\end{align}
where $\kappa$, $A$, and $E(\infty)$ are fitting parameters. We employ Eq.~(\ref{eq:extrapolation}) for all one-dimensional calculations in this work. The horizontal dashed line indicates the extrapolated energy $E(\infty)$ obtained from the dilated lattice results.
The lower panel of Fig.~\ref{fig:3} compares the ground state wave function in $\bm{r}$-space (dashed line) and $\bm{s}$-space (solid line). Due to lattice spacing enhancement at large radii, the wave function depicted in $\bm{s}$-space decays significantly faster than that in $\bm{r}$-space.
While the central region $r \leq R$ remains essentially unchanged, the asymptotic behavior is compressed from $O(e^{-\kappa r})$ to $O(e^{-\lambda \kappa r})$, where $\kappa$ is the binding momentum. For the same number of lattice points (\textit{e.g.}, $L = 30$), the dilated lattice completely captures the wave function tail and converges to the infinite-volume limit. 
In contrast, the uniform lattice still exhibits substantial finite-volume effects.

For general calculations, we can estimate the required box size to achieve convergence. 
For a uniform lattice, if the wave function extends to a radius $R_c$, the minimal box size needed to eliminate finite-volume effects is $L = 2R_c$. 
After applying the dilated coordinate transformation, the wave function size reduces to $R_{c}^\prime = R + (R_c - R) / \lambda$, implying a minimal box size $L^\prime = 2 R_{c}^\prime$. 
From Fig.~\ref{fig:3}, we observe that the binding energy converges at approximately $L/a \approx 100$ for the uniform lattice (corresponding to $R_c / a \approx 50$). 
With dilation $(R=10~\mathrm{fm}, \lambda=3)$, we obtain $R_{c}^\prime/a \approx 23.33$, yielding an estimated minimal box size $L_{c}^\prime/a = 2R_{c}^\prime/a \approx 46$. 
The actual minimal box size determined from Fig.~\ref{fig:3} is $L \approx 30$, which is even smaller than this simple estimate.

\subsection{Three-dimensional two-boson system}
\begin{figure}[h]
\centering
\includegraphics[width=0.45\textwidth]{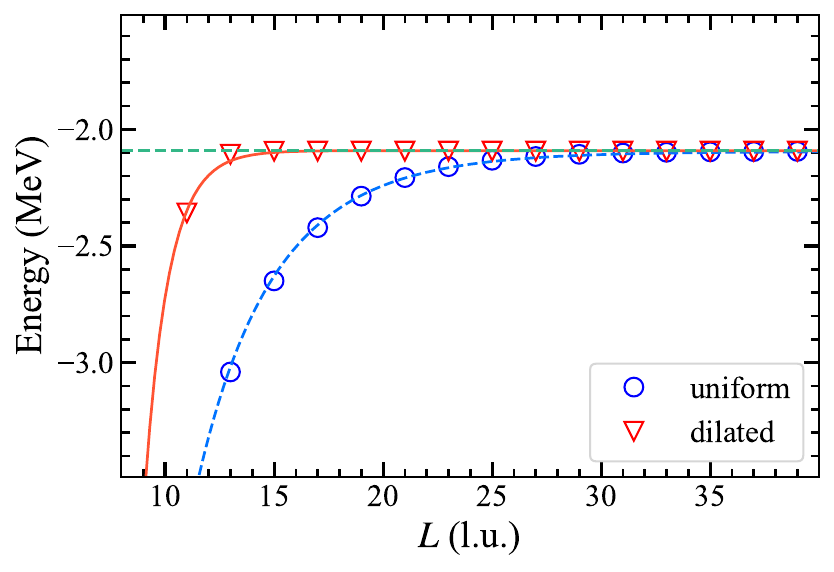}
\caption{Binding energy of three-dimensional two-boson system calculated with uniform (circles) and dilated (triangles) lattices.
Dilation is performed for $R = 8$~fm with the asymptotic ratio $\lambda = 5$.
Lines denote extrapolation function Eq.~(\ref{eq:extrapolation3d}) fitted to the numerical results.
The horizontal dashed line represents infinite-volume limit for dilated lattice.}
\label{fig:10}       
\end{figure}

The above discussion of one-dimensional systems extends readily to three-dimensional calculations. 
We first benchmark the three-dimensional case using two bosons interacting through the same Gaussian potential Eq.~(\ref{eq:GaussianPot}) employed in the one-dimensional case. 
Adopting parameters $C = -40$~MeV and $\sigma = 1.5$~fm, we mimic the strength and range of a typical nucleon-nucleon potential. 
The interaction strength is adjusted to approximately reproduce the deuteron binding energy. 
With the lattice spacing fixed at $a=0.99$~fm, calculations are performed for box sizes $L$ ranging from $10$ to $40$. 
For this and all three-dimensional calculations, we adopt the extrapolation ansatz
\begin{align}
 &E(L) = E(\infty) + \frac{A}{L} \exp(-\kappa L),
\label{eq:extrapolation3d}
\end{align}
with fitting parameters $A$, $\kappa$, and $E(\infty)$. 
Dilated lattice results presented in Fig.~\ref{fig:10} were obtained with dilation parameters $R = 8$~fm and $\lambda = 5$. 
The dilated lattice demonstrates accelerated convergence toward the correct infinite-volume limit $E=-2.09$~MeV, confirming the effectiveness of our method.

\subsection{Deuteron from pionless EFT}
\begin{figure}[h]
\centering
\includegraphics[width=0.45\textwidth]{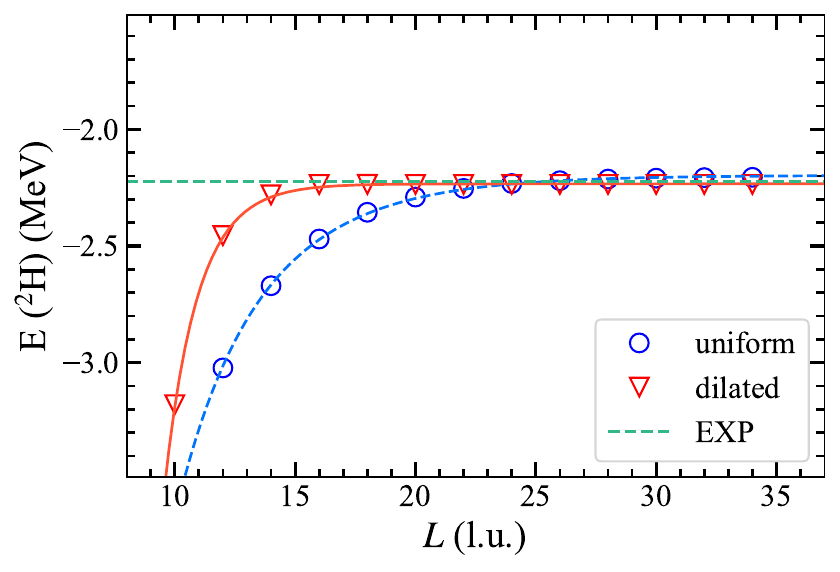}
\caption{Deuteron binding energy calculated with uniform (cicles) and dilated (triangles) lattices versus box size $L$. 
Dilation is performed for $R = 10$~fm with the asymptotic ratio $\lambda = 5$.
Horizontal line denotes the experimental value.}
\label{fig:4}       
\end{figure}

Next we consider realistic nuclear systems, starting with the simplest bound nuclear state, \textit{i.e.}, the deuteron. 
The experimentally measured deuteron binding energy is $-2.224$~MeV~\cite{chupp1961binding}. 
For this verification study, we employ the nuclear force from leading-order (LO) pionless effective field theory,
\begin{align}  
   V_{Q^0}=B_1+B_2(\bm{\sigma_1\cdot\sigma_2}),
   \label{eq:EFTPionless}
\end{align}
where $B_1=-596~\mathrm{MeV}$ and $B_2=-36.4~\mathrm{MeV}$ are low-energy constants (LECs) determined by fitting to the S-wave nucleon-nucleon scattering lengths, following the method in Refs.~\cite{lu2016precise,epelbaum2005two,bogner2010low}. 
Here $\bm{\sigma}_{1,2}$ denote the spin Pauli matrices. 
We fix the lattice spacing to $a = 0.99$~fm and use dilation parameters $\lambda=5$ and $R=10~\mathrm{fm}$. 
The resulting binding energies are presented in Fig.~\ref{fig:4}. 
The horizontal dashed line indicates the experimental value $E({}^{2}\mathrm{H}) = -2.224$~MeV.
We again observe significantly faster convergence using the dilated lattice. 
Both uniform and dilated lattice results converge to the experimental value as $L \rightarrow \infty$, but the dilated lattice prediction is approximately $0.03$ MeV lower than that from the uniform lattice. 
This difference originates from residual lattice artifacts due to the singularity of the EFT potential Eq.~(\ref{eq:EFTPionless}) and can be eliminated by applying an additional soft regulator as in Ref.~\cite{PRL128-242501}.

\subsection{Three-dimensional two-boson system with attractive Coulomb interaction}

Long-range interactions, such as pion-exchange potentials and the Coulomb force, play essential roles in nuclear effective field theory (EFT).
The range of these forces is governed by the mass of the exchanged bosons.
While pion exchange determines the tail of the nuclear force with a range of approximately $1$~fm, the Coulomb force, mediated by the massless photon, possesses an infinite range.
Consequently, the finite-volume effects arising from the Coulomb force present a significant challenge for lattice simulations of nuclei.
Finite-volume corrections for the Coulomb force have been derived both perturbatively~\cite{PRD90-054503, PRD90-074511, Science347-1452, EPJA57-26, PRC103-064611, PRD103-094520} and non-perturbatively~\cite{PRL131-212502}.
Here we examine the interplay between the Coulomb force and the dilated coordinate method.
As for Coulomb force there exists no sharp radius cutoff $R$ beyond which the interaction vanishes completely, the finite volume behavior would be different from the short-range interactions.

We consider energy spectrum of two-bosons interacting exclusively via an attractive Coulomb force,
\begin{equation}
    V(\bm{r}) = -\frac{C_{\mathrm{coul}}}{r},
    \label{eq:coulumb}
\end{equation}
where $C_{\mathrm{coul}}=0.27$ denotes the Coulomb interaction strength.
The energy spectrum can be solved analytically in the continuum and has the familiar form:
\begin{equation}
    E = -\frac{\mu C_{\rm coul}^2}{2n^2},
    \label{eq:-1/n2}
\end{equation}
where $\mu$ is the reduced mass and $n$ is the principle quantum number. 
For lattice calculations, we fix the lattice spacing to $a = 0.99$~fm and the box size to $L/a=63$. 
The finite volume effects would become severe for highly excited states close to the threshold, resulting in deviations from the analytical formula Eq.~(\ref{eq:-1/n2}).
We check whether the dilated lattice help eliminate these unphysical contamination. 
For Coulomb potential we adopt dilation parameters $\lambda=5$ and $\mathrm{R}=15~\mathrm{fm}$.
The singularity of Eq.~(\ref{eq:coulumb}) at  $r=0$ is regularized by introducing a single contact term with its strength $V_{\rm cont}$ adjusted to reproduce the ratio of the ground state and first excited energies $E_1 / E_2 = 1/4$ for both lattices.
We found $V_{\rm cont} = -537.4$~MeV for uniform lattice and $V_{\rm cont} = -533.5$~MeV for dilated lattice, respectively.
\begin{figure}[hbt!]
\centering
\includegraphics[width=0.45\textwidth]{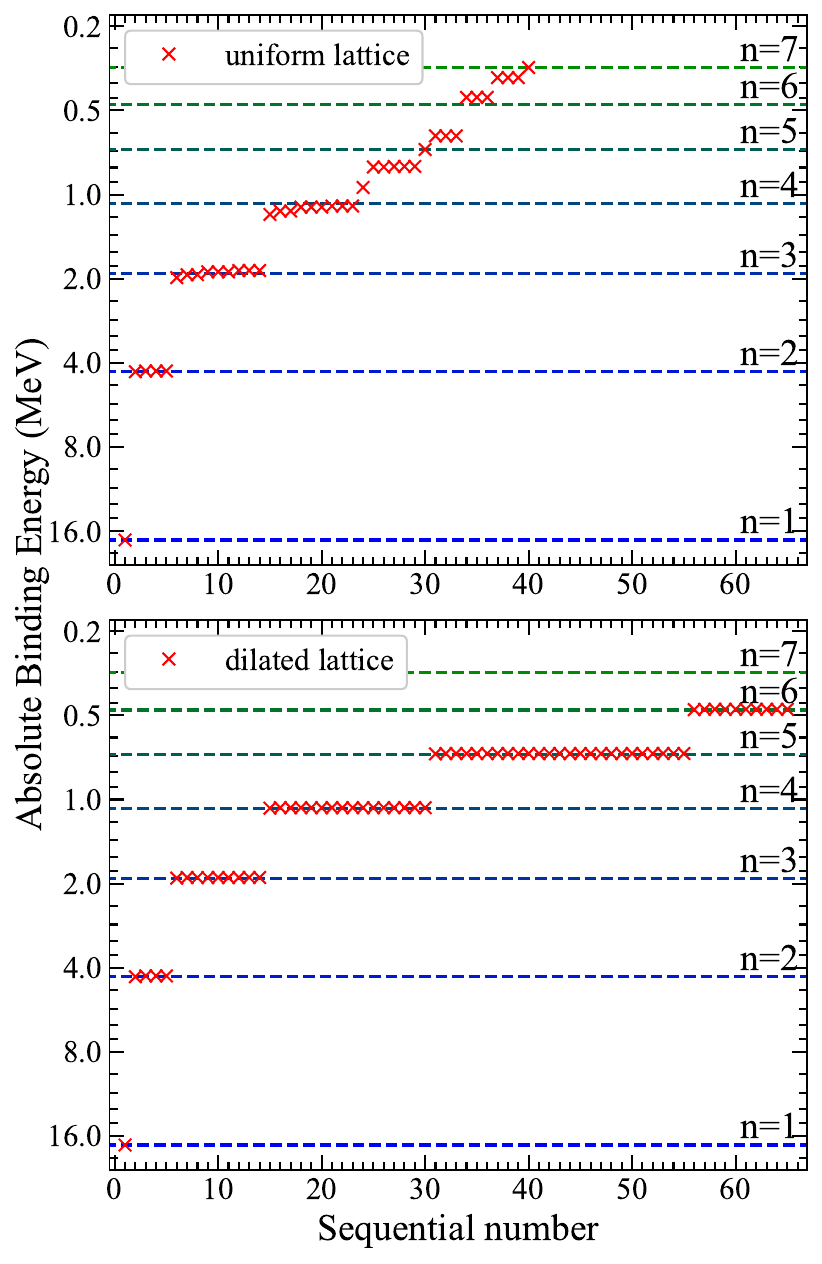}
\caption{Lowest energy levels for an attractive Coulomb potential between two bosons calculated with uniform (Upper) and dilated (Lower) lattices. 
Rightmost numbers $n$ denotes the principal quantum number, colored dashed lines represent exact solutions proportional to $1/n^2$.}
\label{fig:5}
\end{figure}

The upper panel of Fig.~\ref{fig:5} displays the lowest energy levels computed using a uniform lattice.
To improve visibility of highly excited states, we employ a logarithmic scale and present absolute values of the binding energies.
While the ground and first excited states satisfy Eq.~(\ref{eq:-1/n2}) by construction, the second excited state approximately clusters around the corresponding analytical solution.
For the lowest few energy levels ($1 \leq n \leq 3$), the degeneracy pattern follows the expected $n^2$-fold multiplicity.
However, at higher energies both the excitation energies and degeneracy deviate increasingly from the theoretical solution, indicating progressively stronger finite-volume effects.

The lower panel of Fig.~\ref{fig:5} shows the lowest energy levels computed with the dilated lattice. 
Dilated coordinates effectively mitigate finite-volume effects for excited states. 
The results achieve the perfect $n^2$-fold degeneracy for $n \leq 5$ energy levels.
Moreover, all energy levels show remarkable precision, even for highly excited states near the threshold.
\begin{figure}[hbt!]
\centering
\includegraphics[width=0.45\textwidth]{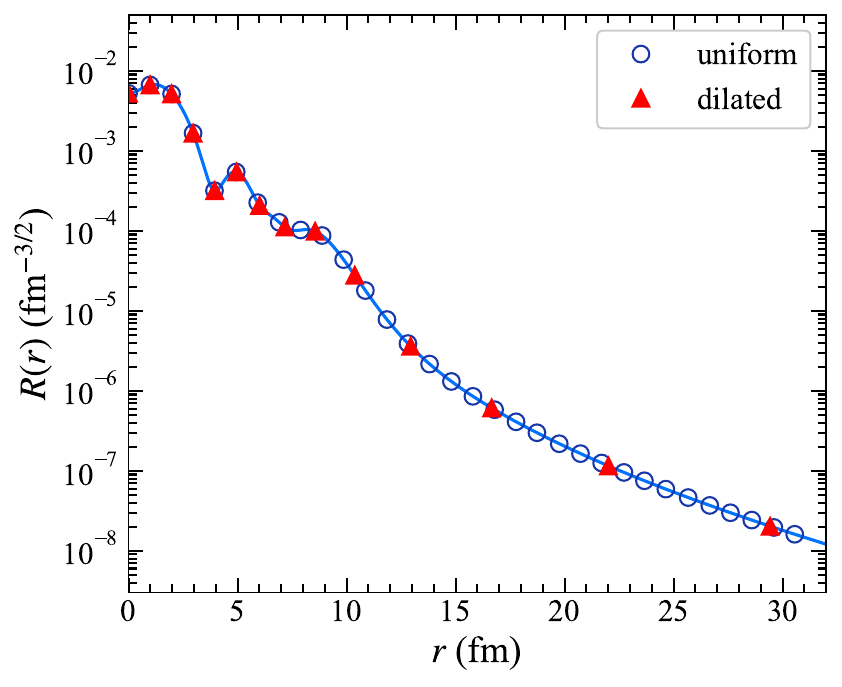}
\caption{Radial wave function $R(r)$ of a 
$n=5$ state in an attractive Coulomb potential. Open circles and full triangles indicate numerical results obtained with uniform and dilated lattices, respectively.}
\label{fig:5-1}
\end{figure}

To examine the effect of dilation on wave functions, Fig.~\ref{fig:5-1} plots the radial component of an $n=5$ wave function versus radial distance $r$.
For comparison, lattice results are shown for both uniform (circles) and dilated (triangles) lattices.
For the dilated lattice case, we first solve for the wave function in $\bm{s}$-space, then transform it to $\bm{r}$-space using Eq.~(\ref{eq:dialteFunction}).
This transformation significantly enlarges the lattice spacing at large $r$.
To visualize the long tail of the wave function more clearly, we use a logarithm scale.
We found that both lattice wave functions coincide for small radius.
While the uniform lattice employs numerous lattice points to resolve the slowly decaying tail, the dilated lattice achieves comparable precision in this region using far fewer points.
This efficiency arises because the wave function becomes sufficiently smooth in the tail region, allowing accurate representation with larger lattice spacings.

\subsection{One-dimensional three-boson system}
\begin{figure}[h]
\centering
\includegraphics[width=0.47\textwidth]{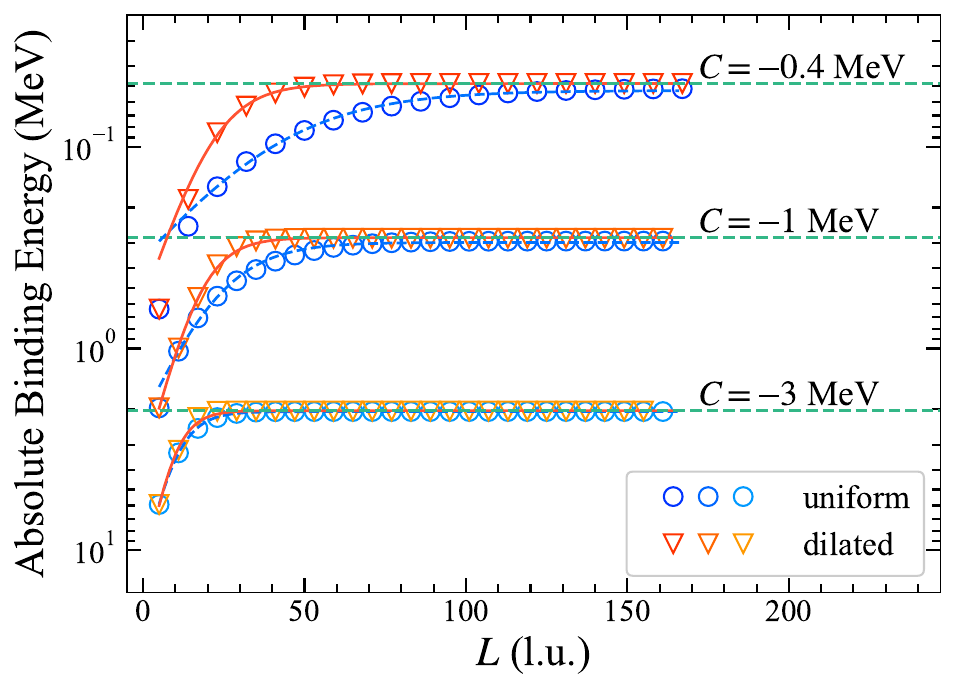}
\caption{Binding energies of one-dimensional three-boson system for different coupling strength $C$ calculated with uniform (circles) and dilated (triangles) lattices. Dilation is performed for $R = 10$~fm and asymptotic ratio $\lambda = 3$.}
\label{fig:9}       
\end{figure}

The computational cost increases more rapidly for the three-body case, thus we are mostly confined to small box size and demands more efficient techniques for eliminating the finite volume effects.
Here we first examine the dilated coordinate method with three identical one-dimensional bosons interacting via the Gaussian potential Eq.~(\ref{eq:GaussianPot}). 
We fix the lattice spacing to $a=0.99$~fm and the interaction range to $\sigma = 1.5$~fm.
To examine the dependence of the results on the strength of the interaction, we compare three different values of $C$.
We take $C=-3,\ -1,\ -0.4$~MeV and repeat the calculations.
The resulting binding energies vary by one order of magnitude and are depicted in Fig.~\ref{fig:9}.
The results with the dilated lattice are computed with the dilation radius $R=10$~fm and ratio $\lambda = 3$.
The dilated lattice gives extrapolated energies $E=-2.04,-0.28,-0.05$~MeV for interaction strength $C=-3,\ -1,\ -0.4$~MeV, respectively. 
For visually clearness, we plot the absolute values of binding energies in logarithmic scale.
We find that without dilation, the shallowly bound state converges much slower than the deep bound state. 
The acceleration of convergence is significantly observed for all situations. 
Note that the results from both lattices are consistent with each other after extrapolation.

\subsection{Three-dimensional three-boson system}
Finally, we investigate a three-boson system in three dimensions.
Such calculations face challenges analogous to those in triton systems within full NLEFT frameworks.
We consider three identical bosons interacting through the pairwise Gaussian potential given by Eq.~(\ref{eq:GaussianPot}).
Parameters are fixed to lattice spacing $a = 0.99$~fm, interaction strength $C = -55$~MeV, and range $\sigma = 1.0$~fm.
The dilation transformation employs $R = 10$~fm and $\lambda = 2.5$.
\begin{figure}[h]
\centering
\includegraphics[width=0.45\textwidth]{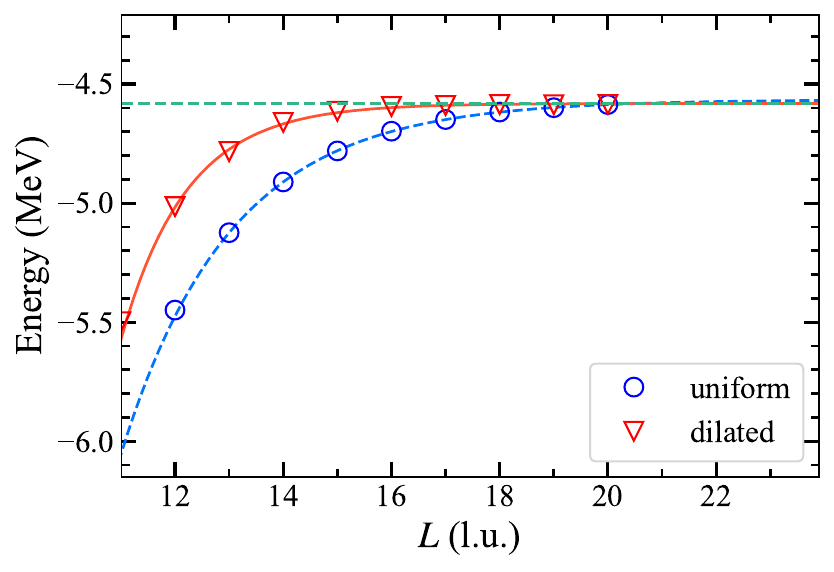}
\caption{Binding energy for the three-boson system in three dimensions computed with uniform (circles) and dilated (triangles) lattices.
Dilation is performed for $R = 10$~fm and asymptotic ratio $\lambda = 2.5$.
Lines show fits of extrapolation function Eq.~(\ref{eq:extrapolation3d}) to numerical results.
The horizontal dashed line indicates the infinite-volume limit for the dilated lattice.}
\label{fig:7}       
\end{figure}
Fig.~\ref{fig:7} presents the resulting energies as functions of box size $L$.
The horizontal dashed line indicates the extrapolated value $E(\infty)$ for the dilated lattice results.
Convergence requires $L/a = 20$ for the uniform lattice, while the dilated lattice achieves convergence at the smaller $L/a \approx 16$.
As for three-dimensional three-body calculations the computational cost scales as $O(L^6)$, we expect a $(20/16)^6 \approx 4$ times acceleration.
The extrapolated energies are $E=-4.565$~MeV and $E=-4.581$~MeV for uniform and dilated lattice, respectively. 
Difference between converged values is approximately 0.0165~MeV, representing $0.36\%$ of the total binding energy.
This residual difference can be further minimized through optimization of the dilation parameters.

\section{conclusion and outlook}

Lattice simulations have found broad applications across diverse fields from computational graphics to finite element analysis. Adaptive mesh refinement (AMR) offers an efficient strategy for reducing computational demands by employing coarse lattices as default and introducing finer resolutions only where necessary. This approach typically lowers computational costs by several orders of magnitude while preserving accuracy, proving particularly valuable for multiscale systems like effective field theories (EFT) where scale separation is intrinsic by definition.

We implement the dilated coordinate method, a specialized AMR technique, in nuclear lattice effective field theory (NLEFT). Traditional NLEFT solves nuclear many-body problems through direct matrix diagonalization or quantum Monte Carlo methods on cubic lattices. For light nuclei, exact diagonalization is feasible but becomes prohibitively expensive as box size $L$ increases. In three-body systems, the wave function dimension scales as $O(L^6)$, restricting typical simulations of triton to $L \leq 16$. This limitation induces significant finite-volume effects, especially for shallow bound states like the ${}^{3}\mathrm{H}$ ground state, necessitating infinite-volume extrapolations. The lattice spacing $a$, fixed by momentum cutoff in NLEFT, creates a fundamental restriction for spatial resolution and physical box size.
Our dilated coordinate method resolves this constraint by introducing variable coordinate-dependent lattice spacings through analytic coordinate transformations. This approach effectively extends the simulation volume without additional computational overhead or accuracy degradation. Numerical demonstrations using one- and three-dimensional toy models confirm the efficacy of our method.
The applications to realistic nuclear forces are straightforward.

This approach significantly enhances {\it ab initio} nuclear calculations, where medium-mass and heavy nucleus predictions depend sensitively on precise few-body inputs. Beyond light nuclei, our method addresses systems where finite-volume effects pose fundamental limitations. For example, neutron-rich nuclei near the dripline exhibit extended valence nucleon wave functions relative to core sizes, motivating halo EFT formulations that explicitly separate these degrees of freedom. Our technique provides significantly enlarged computational volumes for such systems.
Similarly, light $\Lambda$ hypernuclei with small separation energies become accessible targets~\cite{EPJA60-215}. 
More broadly, this framework offers solutions for finite-volume dominated scenarios across computational physics where scale separation governs system behavior.

Here we only benchmarked the method for few-body systems with particle number $A \leq 3$.
For many-body systems, the direct diagonalization method is no longer available, necessitating the use of Monte Carlo simulations.
One advantage of the dilated coordinate method is that it compresses the system wave function while actual calculations utilize the same standard cubic lattice with a transformed Hamiltonian.
Consequently, the computational complexity scales identically with the volume of the system and the number of particles.
Transforming lattice interactions may be challenging.
However, for sufficiently short-range interactions and a large dilation radius $R$, it should be feasible to transform only the kinetic energy term without compromising the accuracy.

The method's capacity to simulate large effective volumes also facilitates nuclear reaction studies when combined with multichannel Monte Carlo approaches like adiabatic projection~\cite{EPJA52-174, PRC92-054612, PRC90-064001, EPJA49-151, Nature528-111}.
In such calculations, short-range effective interactions among nuclear clusters, such as $^4$He, are extracted using Monte Carlo simulations within a small volume for computational efficiency. 
These interactions are then applied on an extraordinarily large lattice to compute scattering and reaction cross sections.
Employing such a dual-lattice approach could introduce additional numerical uncertainties, making it desirable to perform the calculation on a single lattice.
The challenge lies in the requirement for a dense lattice to precisely describe the intrinsic structures of the clusters, while low-energy scattering and reactions of astrophysical interest, typically below several MeV, involve much larger wavelengths for relative cluster motion.
This scale separation motivates the application of the dilated coordinate method.
Taking $\alpha$-$\alpha$ scattering as an example, one can employ a small lattice spacing near the origin and a sparse lattice when the clusters are well separated.
Since effective cluster interactions vanish completely at large separations, leaving only the known kinetic energy term,  resolving the intrinsic structures of the clusters is thus unnecessary in this region.
Applications of our method to astrophysical reactions, such as $p(n,\gamma)d$, using adiabatic projection or similar techniques are possible and actively being pursued.

\color{black}

\section{Acknowledgement}

We thank members of the Nuclear Lattice Effective Field Theory Collaboration for insightful discussions.
This work has been supported by NSAF No. U2330401 and National Natural Science Foundation of China with Grant Nos. ~12547105, ~12275259, ~12375073.

%

\end{document}